\pdfoutput=1

\documentclass[11pt,twoside,a4paper,cmspaper,final,collab]{cms-tdr}
\def\svnVersion{314081}\def\cmsCernNo{2013-037}\def\cmsCernDate{\today}\def\cmsMessage{Published in Physics Letters B as \href{http://dx.doi.org/10.1016/j.physletb.2015.10.067}{\doi{10.1016/j.physletb.2015.10.067.}}}

\begin{document}\cmsNoteHeader{HIG-13-010}

\hyphenation{had-ron-i-za-tion}
\hyphenation{cal-or-i-me-ter}
\hyphenation{de-vices}
\RCS$HeadURL: svn+ssh://pakhotin@svn.cern.ch/reps/tdr2/papers/HIG-13-010/trunk/HIG-13-010.tex $
\RCS$Id: HIG-13-010.tex 273795 2015-01-14 18:35:55Z pakhotin $
\newlength\cmsFigWidth
\ifthenelse{\boolean{cms@external}}{\setlength\cmsFigWidth{0.98\columnwidth}}{\setlength\cmsFigWidth{0.74\textwidth}}
\ifthenelse{\boolean{cms@external}}{\providecommand{\cmsLeft}{top}}{\providecommand{\cmsLeft}{left}}
\ifthenelse{\boolean{cms@external}}{\providecommand{\cmsRight}{bottom}}{\providecommand{\cmsRight}{right}}
\hyphenation{off-line}
\cmsNoteHeader{HIG-13-010}
\title{A search for pair production of new light bosons decaying into muons}

\date{\today}

\abstract{
A search for the pair production of new light bosons, each decaying into a pair of muons, is performed with the CMS experiment at the LHC, using a dataset corresponding to an integrated luminosity of 20.7\fbinv collected in proton-proton collisions at center-of-mass energy of $\sqrt{s} = 8$\TeV. No excess is observed in the data relative to standard model background expectation and a model independent upper limit on the product of the cross section, branching fraction, and acceptance is derived. The results are compared with two benchmark models, the first one in the context of the next-to-minimal supersymmetric standard model, and the second one in scenarios containing a hidden sector, including those predicting a nonnegligible light boson lifetime.
}

\hypersetup{%
pdfauthor={CMS Collaboration},%
pdftitle={A search for pair production of new light bosons decaying into muons},%
pdfsubject={CMS},%
pdfkeywords={LHC, CMS, physics, muon, Higgs, supersymmetry, NMSSM, dark SUSY, hidden sectors, dark matter}}

\maketitle

\section{Introduction}

In July 2012 the ATLAS and CMS collaborations at the CERN LHC announced the discovery of a particle~\cite{:2012gk, :2012gu,Chatrchyan:2013lba} with properties consistent with the standard model (SM) Higgs boson~\cite{Englert:1964et, Higgs:1964ia, Higgs:1964pj, Guralnik:1964eu}. Direct measurements of the production and decay rates of the new particle, using SM decay channels, have so far played a key role in determining whether or not it is indeed consistent with the SM predictions. However, substantially increasing the precision of these measurements will require further data. Searches for Higgs bosons through production mechanisms not predicted by the SM, or decay modes involving particles not included in the SM, provide a complementary approach and have the advantage of probing specific types of new physics models with the existing data.

This letter presents a search for the pair production of new light bosons (denoted as `$\Pa$') decaying to pairs of isolated, oppositely charged muons (dimuons). One production mechanism for these new bosons is in the decay chain of a Higgs boson $\Ph$, which can be SM-like or not: $\Ph \to 2\Pa + \mathrm{X} \to 4 \mu + \mathrm{X}$, where $\mathrm{X}$ denotes possible additional particles from cascade decays of the Higgs boson. A range of new physics scenarios predict this decay topology, including the next-to-minimal supersymmetric standard model (NMSSM)~\cite{Ellwanger:1997pa} and models with hidden (or dark) sectors~\cite{ArkaniHamed:2008qn, Baumgart:2009tn, Falkowski:2010cm}.

The NMSSM is an extension of the minimal supersymmetric standard model (MSSM)~\cite{Nilles:1983ge,Chung:2003fi} that includes an additional gauge singlet field. It resolves the so-called $\mu$ problem~\cite{1984PhLB..138..150K} and significantly reduces the amount of fine tuning required in the MSSM~\cite{Casas:2003jx}. The NMSSM Higgs sector consists of three CP-even neutral Higgs bosons $\Ph_{1,2,3}$, two CP-odd neutral Higgs bosons $\Pa_{1,2}$ and a pair of charged Higgs bosons $\Ph^{\pm}$. The $\Ph_1$ and $\Ph_2$ can decay via $\Ph_{1,2} \to 2\Pa_1$, where either the $\Ph_1$ or $\Ph_2$ can be the boson observed at 125\GeV. The $\Pa_1$ boson can be light and couple weakly to SM particles with a coupling to fermions proportional to the fermion mass. Therefore it can have a substantial branching fraction $\mathcal{B}(\Pa_1 \to \PGmp \PGmm)$ if its mass is within the range $2m_{\mu}<m_{\Pa_1}<2m_{\tau}$~\cite{Belyaev:2010ka, Dermisek:2010mg} (benchmark model 1 in this letter). A search for final states containing muon pairs provides sensitivity to models of this form.

Supersymmetry (SUSY) models with dark sectors (dark SUSY) offer an explanation for the excess in the ratio of the positron flux to the combined flux of positrons and electrons observed by the satellite experiments~\cite{FermiLAT:2011ab, Accardo:2014lma,PhysRevLett.111.081102} in primary cosmic rays as well as predict cold dark matter with a scale of $\mathcal{O}(1\TeV)$. A simple realization of these models includes a new $\text{U}(1)_{\mathrm{D}}$ symmetry (the subscript ``D" stands for ``Dark") which is broken and gives rise to massive dark photons (denoted as $\gamma_{\mathrm{D}}$). Kinetic mixing of the new $\text{U}(1)_{\mathrm{D}}$ with the SM hypercharge $\text{U}(1)_{\text{Y}}$ provides a small mixing between $\gamma_{\mathrm{D}}$ and the SM photon which allows $\gamma_{\mathrm{D}}$ to decay to SM particles~\cite{Holdom:1985ag}. Depending on the value $\varepsilon$ of the kinetic mixing, the $\gamma_{\mathrm{D}}$ may also be long-lived. The lack of an antiproton to proton ratio excess of the magnitude similar to the positron excess in the measurements of the cosmic ray spectrum constrains the mass of $\gamma_{\mathrm{D}}$ to be less than twice the mass of the proton~\cite{Adriani:2010rc}. If the hidden sector directly or indirectly interacts with the Higgs field, a number of possible scenarios may be realized. One such scenario, denoted in this letter as benchmark model 2, is a model of SUSY where the SM-like Higgs boson can decay via $\Ph \to 2 \mathrm{n}_1$, where $\mathrm{n}_1$ is the lightest neutralino in the visible (as opposed to hidden) part of the SUSY spectrum. The $\mathrm{n}_1$ can decay via $\mathrm{n}_1 \rightarrow \mathrm{n}_{\mathrm{D}} + \gamma_{\mathrm{D}}$, where $\mathrm{n}_{\mathrm{D}}$ is a dark neutralino that escapes detection. Assuming that $\gamma_{\mathrm{D}}$ can only decay to SM particles, the branching fraction $\mathcal{B}(\gamma_{\mathrm{D}} \to \PGmp \PGmm)$ can be as large as $45\%$, depending on the mass of $\gamma_{\mathrm{D}}$~\cite{Falkowski:2010cm}.

Previous searches for pair production of new light bosons decaying into dimuons were performed at the Tevatron~\cite{Abazov:2009yi} and the LHC~\cite{Aad:2012kw,Chatrchyan:2012cg}. Searches for associated production of the light CP-odd scalar bosons have been performed at $\Pep\Pem$ colliders~\cite{Love:2008aa, Aubert:2009cp} and the Tevatron~\cite{Aaltonen:2011aj}. Direct $\Pa_1$ production has been studied at the LHC~\cite{Chatrchyan:2012am}, but this is heavily suppressed by the typically very weak couplings of the new bosons to SM particles. The constraints on the allowed NMSSM parameter space are driven by the measurements of relic density by WMAP~\cite{Hinshaw:2012aka} and more recently by PLANCK~\cite{Ade:2013zuv}, while specifically for the Higgs sector the most relevant measurements come from LEP~\cite{Abbiendi:2002qp, Abbiendi:2002in, Abbiendi:2004ww, Abdallah:2004wy, Schael:2010aw, Schael:2006cr}, LHC measurements of the SM-like Higgs properties, and direct searches for $\Ph \to \Pa\Pa $~\cite{Chatrchyan:2012cg}. In the framework of dark SUSY, experimental searches for dark photons have focused on their production at the end of SUSY cascades at the Tevatron~\cite{Abazov:2009hn,Abazov:2010uc,Aaltonen:2012pu} and the LHC~\cite{Chatrchyan:2011hr, Aad:2014yea}. Searches at a range of low energy $\Pep\Pem$ colliders ( KLOE~\cite{Babusci:2014sta}, BaBar~\cite{Lees:2014xha}), heavy-ion colliders (PHENIX~\cite{Adare:2014mgk}), fixed-target experiments (APEX~\cite{Abrahamyan:2011gv}, A1 at MAMI~\cite{Merkel:2014avp}, HADES~\cite{Agakishiev:2013fwl}), as well as cosmological measurements~\cite{Fradette:2014sza,2012arXiv1201.2683D,Dreiner:2013mua} and others~\cite{Essig:2013lka,Blumlein:2011mv,Essig:2010gu,Batell:2009di,Gninenko:2012eq} provide constraints on complementary regions of the available parameter space.

Results are presented in this Letter in the context of the two benchmark scenarios discussed earlier, one in the context of NMSSM and another one in the framework of dark SUSY scenarios. However, the search has been designed to be independent of the details of these two specific models, and the results can be interpreted in the context of other models predicting the production of the same final states. Compared to the previous version~\cite{Chatrchyan:2012cg}, the present analysis has been redesigned to be sensitive to signatures with the intermediate bosons traversing a nonnegligible distance before decaying into a pair of muons. Such signatures can be realized in dark SUSY models if the mixing of the dark photon with its SM counterpart is sufficiently weak. In addition, the present analysis uses a dataset four times larger than the previous analysis, and at a higher centre-of-mass energy, further extending the reach for signatures with prompt muons.

\section{The CMS detector}

This search is based on a data sample corresponding to an integrated luminosity of 20.7\fbinv of proton-proton collisions at a center-of-mass energy $\sqrt{s} = 8\TeV$, recorded by the CMS detector in 2012. The central feature of the CMS apparatus is a superconducting solenoid of 6\unit{m} internal diameter, providing a magnetic field of 3.8\unit{T}. Within the solenoid volume are a silicon pixel and strip tracker, a lead tungstate crystal electromagnetic calorimeter, and a brass and scintillator hadron calorimeter, each composed of a barrel and two endcap sections. Muons are measured in gas-ionization detectors embedded in the steel flux-return yoke outside the solenoid. Extensive forward calorimetry complements the coverage provided by the barrel and endcap detectors. Muons are measured in the pseudorapidity range $\abs{\eta}< 2.4$, with detection planes made using three technologies: drift tubes, cathode strip chambers, and resistive plate chambers. Matching muon candidates to tracks measured in the silicon tracker results in an accurate measurement of the transverse momentum (\pt). As an example, for muons with $\pt < 10$\GeV the relative \pt\ resolution is found to be 0.8\%--3.0\% (depending on $\abs{\eta}$) and for muons with $20 <\pt < 100$\GeV it is 1.3--2.0\% in the barrel and better than 6\% in the endcaps~\cite{CMS_Muon_Reco}. A more detailed description of the CMS detector, together with definitions of the coordinate system used and the relevant kinematic variables, can be found in~\cite{Chatrchyan:2008aa}.

\section{Data selection}

The data were collected with an online trigger selecting events containing at least two muon candidates, one with $\pt > 17$\GeV and another with $\pt > 8$\GeV. In this analysis offline muon candidates are defined as particle-flow (PF) muons~\cite{CMS_Muon_Reco}. The PF reconstruction algorithm combines information from all CMS subdetectors to identify and reconstruct individual particles, such as electrons, photons, hadrons or muons.

Events are further selected by requiring at least four offline muon candidates with $\pt > 8$\GeV and $\abs{\eta} < 2.4$ that form two oppositely charged pairs. At least one of these muons must additionally satisfy the requirement of $\pt > 17$\GeV and $\abs{\eta} < 0.9$, which ensures that the trigger efficiency is high and independent of the event topology, including effects related to overlaps of nearby muon trajectories. Tracks associated with a pair of opposite-charge muon candidates are fit for a common vertex using a Kalman filter algorithm~\cite{Luchsinger:1992np}. If the vertex is reconstructed, a muon pair is combined into a dimuon system if its invariant mass measured at the common vertex $m_{\PGmp \PGmm} < 5$\GeV and the vertex fit probability $P_v(\PGmp \PGmm) > 1\%$. Muon pairs failing these requirements are still retained for the analysis if at the point of closest approach of the two trajectories they are within $\Delta R (\PGmp, \PGmm) = \sqrt{(\eta_{\PGmp} - \eta_{\PGmm})^2 + (\phi_{\PGmp} - \phi_{\PGmm})^2} < 0.01$, where $\phi_{\mu^\pm}$ are the azimuthal angles in radians. This recovery step is designed to compensate for the reduced efficiency of the vertex selection for dimuons in which the two muon tracks are nearly parallel to each other, therefore a good efficiency is maintained for dimuon masses down to the $2 m_{\mu}$ threshold (0.2114\GeV). For dimuons in which this is the case the point of closest approach is selected as the vertex position, with the additional selection requirement that the distance between the tracks be $\leq$0.5\unit{mm}. The dimuon kinematic variables are measured at the dimuon vertex position. There is no restriction on the number of ungrouped additional muons. Both dimuons are required to have at least one hit in the first layer of the barrel or endcaps of the pixel detector, and this defines an effective ``fiducial'' region. This requirement, along with the muon $\pt$ and $\abs{\eta}$ criteria, ensures high trigger ($>$96\%), reconstruction, and selection efficiencies, with a greatly reduced dependence on the \pt, $\eta$, or opening angle between the muons.

The projected $z$ coordinate of the dimuon system at the point of the closest approach to the beam line ($z_{\mu\mu}$) is reconstructed using the dimuon momentum. The requirement $\abs{z_{1\mu\mu} - z_{2\mu\mu}}<1$\unit{mm} is imposed to ensure that both dimuons are consistent with the same $\Pp\Pp$ interaction; no explicit requirements are made on the impact parameter or the $z$ coordinate at the point of closest approach to the beam line of the individual reconstructed muons to preserve sensitivity to signatures with displaced muons.

To suppress background events in which the muons are produced in the decay of heavy quarks (and thus appear in jets), the dimuons are required to be isolated from other event activity using the criterion $I_{\text{sum}} < 2$\GeV. The isolation parameter $I_{\text{sum}}$ is defined as the scalar sum of the \pt of charged tracks with $\pt > 0.5$\GeV within a cone of size $\Delta R=0.4$ centered on the momentum vector of the dimuon system, excluding the tracks corresponding to the two muon candidates. The tracks used in the calculation of $I_{\text{sum}}$ must also have a $z$ coordinate at the point of closest approach to the beam line that lies within 1\unit{mm} of $z_{\mu\mu}$. The $I_{\text{sum}}$ selection suppresses the contamination from $\bbbar$ production by about a factor of 40, as estimated using a $\bbbar$ enriched control sample with one dimuon recoiling off a jet containing an unpaired muon, while rejecting less than 20\% of events with the signal topology.

The invariant mass $m_{1\mu\mu}$ always refers to the dimuon containing a muon with $\pt > 17$\GeV and $\abs{\eta}<0.9$. For events with both dimuon systems containing such a muon, the assignment of $m_{1\mu\mu}$ and $m_{2\mu\mu}$ is random for compatibility with the background modeling schema described in Section~\ref{sec:background}. The invariant masses of both reconstructed dimuons are required to be compatible within the detector resolution, specifically $\abs{m_{1\mu\mu} - m_{2\mu\mu}} < 0.13\GeV  + 0.065 \,(m_{1\mu\mu}+m_{2\mu\mu})/2$, which defines a diagonal signal region in the plane of the invariant masses of the two dimuons. The numerical parameters in the requirement correspond to at least five times the size of the core resolution in the dimuon mass.

\section{Signal modeling}

The results from this analysis are designed to be model independent, but are also presented in the context of the two benchmark models introduced earlier. NMSSM simulation samples for benchmark model 1 are generated with \PYTHIA~6.4.26~\cite{Sjostrand:2006za}, using MSSM Higgs boson production via gluon fusion $\Pg\Pg \to {\PHz}_\mathrm{MSSM}$, with the Higgs bosons decaying via ${\PHz}_\mathrm{MSSM} \to 2\text{A}^0_\mathrm{MSSM}$. The masses of the MSSM bosons ${\PHz}_\mathrm{MSSM}$ and $\text{A}^0_\mathrm{MSSM}$ are set to the desired values for the $\Ph_1$ mass and $\Pa_1$ mass of the NMSSM bosons, respectively. The mass of ${\PHz}_\mathrm{MSSM}$ is in the range $90 - 150$\GeV (mass below 90\GeV is excluded by LEP~\cite{Schael:2006cr}) and the mass of $\text{A}^0_\mathrm{MSSM}$ is in range $0.25 - 3.55$\GeV. Both $\text{A}^0_\mathrm{MSSM}$ bosons are forced to decay promptly to a pair of muons. Dark SUSY simulation samples for benchmark model 2 are generated with \MADGRAPH~4.5.2~\cite{Alwall:2007st} using SM Higgs boson production via gluon fusion $\Pg\Pg \to \Ph_\mathrm{SM}$, with $m_{\Ph_\mathrm{SM}} = 125$\GeV. The \textsc{Bridge} program~\cite{Meade:2007js} is used to force the Higgs bosons to undergo a non-SM decay to a pair of neutralinos, each of which decays via $\mathrm{n}_1 \to \mathrm{n}_{\mathrm{D}} + \gamma_{\mathrm{D}}$, where $m_{\mathrm{n}_1} = 10$\GeV, $m_{\mathrm{n}_{\mathrm{D}}} = 1$\GeV, which is representative of the type of models considered~\cite{Aad:2014yea}. Dark photons are generated with $m_{\gamma_{\mathrm{D}}}$ in the range 0.25--2.0\GeV and a decay length $c\tau_{\gamma_{\mathrm{D}}}$ in the range of 0--20\unit{mm}. Each of the two dark photons are forced to decay to two muons, while both dark neutralinos escape detection. The narrow-width approximation is imposed by setting the widths of the dark photons to a small value ($10^{-3}$\GeV).

\begin{table*}[t]
\topcaption{Event selection efficiencies $\epsilon_{\text{sim}}(m_{\Ph_1}, m_{\Pa_1})$ and $\epsilon_{\text{sim}}(m_{\gamma_{\mathrm{D}}}, c\tau_{\gamma_{\mathrm{D}}})$, as obtained from simulation, the geometric and kinematic acceptances $\alpha_{\text{gen}}(m_{\Ph_1}, m_{\Pa_1})$ and $\alpha_{\text{gen}}(m_{\gamma_{\mathrm{D}}}, c\tau_{\gamma_{\mathrm{D}}})$, calculated using only generator-level information, and their ratios (with statistical uncertainties), for a few representative NMSSM and dark SUSY benchmark samples. The experimental data-to-simulation scale factor ($\epsilon_{\text{data}}/\epsilon_{\text{sim}}$, described later) is not applied.\label{tab:efficiency_NMSSM_SUSY}}
\begin{center}
\begin{footnotesize}
\begin{tabular}{c |c c c}
\hline
$m_{\Ph_1} \: $[\GeVns{}] & $ 90 $  & $ 125   $ &   $ 125 $ \\
$m_{\Pa_1} \: $[\GeVns{}] &  $ 2   $ & $ 0.5   $  &  $ 3.55   $ \\ \hline
$\epsilon_{\text{sim}} \: [\%]$ & $11.0 \pm 0.1$  &  $21.1 \pm 0.1$  & $17.3 \pm 0.1$ \\
$\alpha_{\text{gen}} \: [\%]$ & $15.9 \pm 0.1$ &   $32.0 \pm 0.1$  & $26.3 \pm 0.1$ \\
$\epsilon_{\text{sim}}/\alpha_{\text{gen}} \: $ &  $0.69 \pm 0.01$  & $0.66 \pm 0.01$  & $0.66 \pm 0.01$ \\
\hline
\multicolumn{4}{c}{} \\
\end{tabular}
\begin{tabular}{c | c c c |  c c c}
\hline
$m_{\gamma_{\mathrm{D}}} \: $[\GeVns{}] & \multicolumn{3}{c|}{0.25}   &   \multicolumn{3}{c}{1.0} \\
$c\tau_{\gamma_{\mathrm{D}}} \: [\text{mm}]$ & $ 0   $ & $ 0.5   $ & $ 2   $  &   $ 0   $ & $ 0.5   $ & $ 2   $ \\
\hline
$\epsilon_{\text{sim}} \: [\%]$  &  $8.85 \pm 0.12$ & $1.76 \pm 0.05$ & $0.23 \pm 0.03$  &  $6.13 \pm 0.23$ & $4.73 \pm 0.07$ &$1.15 \pm 0.04$\\
$\alpha_{\text{gen}} \: [\%]$  &  $14.32 \pm 0.14 $ & $2.7 \pm 0.06$ & $0.31 \pm 0.03$ &  $8.89 \pm 0.28$ & $6.98 \pm 0.09$ &$1.68 \pm 0.05$\\
$\epsilon_{\text{sim}}/\alpha_{\text{gen}} \: $ & $0.62 \pm 0.01$ & $0.65 \pm 0.02$ & $0.74 \pm 0.13$  &$0.69 \pm 0.03$ &$0.68 \pm 0.01$ &$0.68 \pm 0.03$ \\
\hline
\end{tabular}
\end{footnotesize}
\end{center}
\end{table*}

All benchmark samples are generated using the leading-order CTEQ6.6~\cite{Nadolsky:2008zw} set of parton distribution functions (PDF), and are interfaced with \PYTHIA~using the Z2* tune~\cite{Chatrchyan:2011id} for the underlying event activity at the LHC and to simulate jet fragmentation.

The signal samples are processed through a detailed simulation of the CMS detector based on \GEANTfour~\cite{Agostinelli:2002hh} and are reconstructed with the same algorithms used for data. Table~\ref{tab:efficiency_NMSSM_SUSY} shows the event selection efficiencies $\epsilon_{\text{sim}}$ obtained using the simulated benchmark samples for a few representative choices of $(m_{\Ph_1}, m_{\Pa_1})$, and $(m_{\gamma_{\mathrm{D}}}, c\tau_{\gamma_{\mathrm{D}}})$. To provide a simple recipe for future reinterpretations of the results in the context of other models, the variable $\alpha_{\text{gen}}$ is separately defined as the geometric and kinematic acceptance of this analysis calculated using only generator-level information. It is defined by selecting events containing at least four muons with $\pt > 8$\GeV and $\abs{\eta} < 2.4$, with at least one of these muons having $\pt > 17$\GeV and $\abs{\eta} < 0.9$. The new light boson must also decay with transverse decay length $L_{xy} < 4.4$~cm and longitudinal decay length $L_z < 34.5$~cm (both defined in the detector reference frame), to satisfy the ``fiducial'' region of the analysis. Table~\ref{tab:efficiency_NMSSM_SUSY} shows $\alpha_{\text{gen}}$ along with the ratio $\epsilon_{\text{sim}}/\alpha_{\text{gen}}$.

\section{Background estimation}
\label{sec:background}

The SM background for this search is dominated by $\bbbar$ production and has small contributions from the electroweak production of four muons and direct \JPsi pair production. The leading part of the $\bbbar$ contribution is due to $\cPqb$ quark decays that result in a pair of muons, via either the semileptonic decays of both the $\cPqb$ quark and the resulting $\cPqc$ quark, or via resonances, i.e. $\omega$, $\rho$, $\phi$, \JPsi. A smaller contribution comes from events with one genuine dimuon candidate and a second dimuon candidate containing one muon from a semileptonic \cPqb\ quark decay and a charged hadron misidentified as another muon.

Using data control samples, the $\bbbar$ background is modeled as a two-dimensional (2D) template $B_{\bbbar}(m_{1\mu\mu},m_{2\mu\mu})$ in the plane of the invariant masses of the two dimuons. The template describing the 2D probability density function is constructed as a Cartesian product $B_{17}(m_{1\mu\mu}) \, B_{8}(m_{2\mu\mu})$, where the $B_{17}$ and $B_{8}$ templates model the invariant mass distributions for dimuons with and without the requirement that the dimuon contains at least one muon satisfying $\pt > 17$\GeV and $\abs{\eta}<0.9$ respectively. The $B_{17}$ shape is measured using a data sample enriched with $\bbbar$ events containing exactly one dimuon and one additional muon, under the assumption that the decay of one of the $\cPqb$ quarks results in a dimuon pair containing at least one muon with $\pt > 17$\GeV and $\abs{\eta}<0.9$, while the other $\cPqb$ quark decays semileptonically resulting in the additional muon with $\pt > 8$\GeV. For the $B_{8}$ shape, a similar sample and procedure is used but the dimuon is required to have both prongs with $\pt > 8$\GeV, while the additional muon must have $\pt > 17$\GeV and $\abs{\eta}<0.9$. The two templates are required as the shape of the dimuon invariant mass distribution depends on the \pt thresholds used to select the muons and whether the muons are restricted to the central $(\abs{\eta} < 0.9)$ region or can be in the full acceptance range ($| \eta | < 2.4)$, as a result of the differences in the momentum resolution of the barrel and endcap regions of the tracker. The $B_{17}$ and $B_{8}$ distributions are fitted with a parametric analytical function using a sum of Bernstein polynomials and Crystal Ball functions~\cite{Oreglia:1980cs} describing resonances. These event samples do not overlap with the sample containing two dimuons that is used for the main analysis and they have negligible contributions from non-$\bbbar$ backgrounds. Once the $B_{\bbbar}(m_{1\mu\mu},m_{2\mu\mu})$ template is constructed, it is used to provide a description of the $\bbbar$ background shape in the signal region. This technique assumes that each $\cPqb$ quark fragments independently and that if the shapes of these distributions are measured using data samples with kinematics very similar to that of the background events then the effects of residual kinematically-induced correlations are small (albeit weakly induced, the shape depends on the $\cPqb$ jet \pt and the \pt of the two $\cPqb$ jets in background $\bbbar$ events tend to be similar). The background template is validated in a region where both dimuons fail the $I_{\text{sum}} <2$\GeV requirement and good agreement with data is observed.

\begin{figure}[tb]
\begin{center}
\includegraphics[width=\cmsFigWidth]{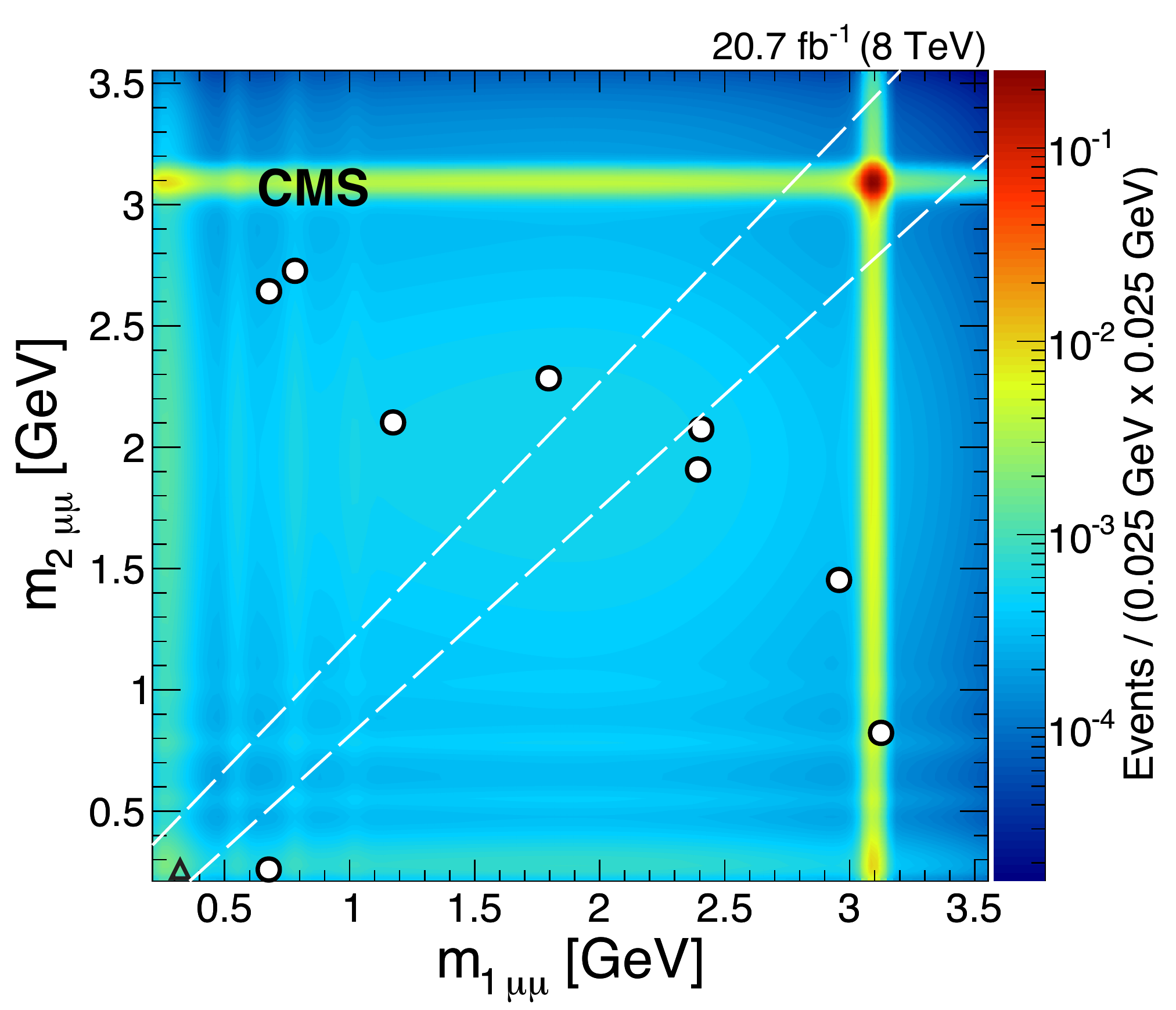}
\end{center}
\caption{Distribution of the invariant masses $m_{1\mu\mu}$ vs. $m_{2\mu\mu}$ for the isolated dimuon events following the application of all constraints except the $m_{1\mu\mu} \simeq m_{2\mu\mu}$ requirement of compatibility within the detector resolution. The compatible diagonal signal region (outlined with dashed lines) contains one data event (triangle) at $m_{1\mu\mu} = 0.33 \GeV$ and $m_{2\mu\mu} = 0.22 \GeV$. There are also nine data events (white circles) which fail the $m_{1\mu\mu} \simeq m_{2\mu\mu}$ compatibility requirement.  The color scale indicates the expected SM background in range $2 m_{\mu} < m_{1\mu\mu}, m_{2\mu\mu} < 2m_{\tau}$. \label{fig:2dtemplate} }
\end{figure}

The data events that satisfy all analysis selections but fail the $m_{1\mu\mu} \simeq m_{2\mu\mu}$ requirement are used to normalize the $B_{\bbbar}(m_{1\mu\mu},m_{2\mu\mu})$ template. This selection yields nine events in the off-diagonal sideband region of the $(m_{1\mu\mu},m_{2\mu\mu})$ plane, leading in the diagonal signal region to an expected rate of $\bbbar$ background events of $2.0 \pm 0.7$. This is essentially $(9 \pm \sqrt{9}) \times 0.18 / 0.82$, where $0.18$ and $0.82$ correspond to the integral of the areas under the background template inside and outside the signal diagonal region, respectively. These nine events in the off-diagonal sidebands of the $(m_{1\mu\mu},m_{2\mu\mu})$ plane are shown as white circles in Fig.~\ref{fig:2dtemplate}.

The contribution from direct \JPsi pair production is estimated using another data control sample. Events are selected with a trigger that requires at least three muon candidates, two of which have a common vertex and an invariant mass consistent with that of the \JPsi particle. Events are further required to contain at least four reconstructed muons with $\pt > 3.5$\GeV, which form dimuon pairs. This control sample does not specifically require that the dimuons satisfy the requirement $I_{\text{sum}}<2$\GeV since $I_{\text{sum}}$ is used to separate the contribution of ``prompt'' and ``nonprompt'' (from $\cPqb$ quark decays) \JPsi in data. Finally, both dimuons are required to have an invariant mass between 2.8 and 3.3\GeV. Following these requirements the data sample consists of events containing prompt and nonprompt \JPsi. To subtract the nonprompt component, two independent methods have been studied: the first one divides the control sample based on the values of the isolation variable $I_{\text{sum}}$ for each of the two dimuons in each event. The number of events in which both dimuons satisfy the requirement $I_{\text{sum}}<2$\GeV is extrapolated from the regions in which at least one of the dimuons fails this requirement. The second approach uses the lifetime of \JPsi candidate, calculated under the hypothesis of it being produced at the beam line, as a discriminating variable. The data distribution is fitted in the isolated region using prompt and nonprompt templates from simulation and nonisolated sideband in data, respectively. Both approaches give consistent results within the associated uncertainties and the results of the isolation-based method are used in the final analysis. There are two mechanisms for the production of prompt double \JPsi events: single- and double-parton scattering (SPS and DPS, respectively), corresponding to whether the two \JPsi mesons are produced from one or two independent parton interactions. The number of prompt events in the control region are further separated into SPS and DPS components using the \JPsi rapidity difference as the discriminating variable. Finally, the data-to-simulation normalization factor and the fraction of SPS and DPS events are extrapolated from the control to the signal region, resulting in a final estimation for the contribution from prompt double \JPsi events of $0.06 \pm 0.03$ events.

The contribution from other SM processes (low mass Drell--Yan production and $\Pp\Pp \to \cPZ/\gamma^* \to 4 \mu$) is estimated with the \textsc{CalcHEP}~3.6.18 generator~\cite{Belyaev:2012qa} using the HEPMDB infrastructure~\cite{hepmdb}, and is found to be $0.15 \pm 0.03$ events in the entire signal region. The combined expected background contribution to the diagonal signal region is $2.2 \pm 0.7$ events. This background is represented by the color scale in Fig.~\ref{fig:2dtemplate}.

\section{Systematic uncertainties}

The selection efficiencies for the offline muon reconstruction, trigger, and dimuon isolation requirements are obtained from simulation, and are corrected with scale factors derived from comparison between data and simulation using $\cPZ \to \mu \mu$ and $\JPsi \to \mu \mu$ samples. The scale factor per event is found to be $\epsilon_{\text{data}}/\epsilon_{\text{sim}} = 0.93 \pm 0.07$ and it accounts for the differences in the efficiency of the trigger, the efficiency of the muon reconstruction and identification for each of the four muon candidates, and the combined efficiency of the isolation requirement for the two dimuon candidates. The estimate accounts for correlations associated with the presence of multiple muons per event. The main systematic uncertainty is the offline muon reconstruction (4.1\%), which includes an uncertainty (1\% per muon) to cover variations of the scale factor as a function of the muon $\pt$ and $\eta$. Other systematic uncertainties include: the uncertainty in  dimuon reconstruction effects related to overlaps of muon trajectories in the tracker and in the muon system (3.5\%), the trigger efficiency (1.5\%), the uncertainty in the efficiency caused by the modeling of the tails in the dimuon invariant mass distribution that arises from the requirement that the two dimuon masses are compatible (1.5\%), and the dimuon isolation (negligible). The uncertainty in the integrated luminosity of the data sample (2.6\%)~\cite{CMS-PAS-LUM-13-001} is also included. All uncertainties quoted above are related to variations in the signal efficiency due to experimental selection and sum up to 6.3\%. The uncertainties related to variations in the signal acceptance due to the model include:  the uncertainties related to the PDFs and the knowledge of the strong coupling constant $\alpha_s$, which are estimated by comparing the PDFs in CTEQ6.6~\cite{Nadolsky:2008zw} with those in NNPDF2.0~\cite{Ball:2010de} and MSTW2008~\cite{Martin:2009iq}, following the PDF4LHC recommendations~\cite{Alekhin:2011sk,Botje:2011sn}. Using the analysis benchmark samples, they are found to be 3\% for the signal acceptance. The variation of the renormalization and factorization scales has a negligible effect. In addition a re-weighting procedure is applied to the Higgs boson \pt\ spectra in the benchmark signal samples to reproduce the NNLO+NNLL prediction~\cite{deFlorian:2012mx} and account for possible changes to the analysis acceptance. Only a weak sensitivity to this is expected and the result of the re-weighting procedure is limited by the statistical uncertainty of the simulated samples (2\%), which is therefore included as a conservative systematic uncertainty on this effect. Thus, the total systematic uncertainty in the signal acceptance and selection efficiency is 7.3\%.

\section{Results}

After the full analysis selection is applied to the data sample, one event is observed in the diagonal signal region, as shown in Fig.~\ref{fig:2dtemplate}. This is consistent with the expected background contribution of 2.2$\pm$0.7 events.

For future reinterpretations of this analysis, the results can be presented as a 95\% confidence level (CL) upper limit on $\sigma(\Pp\Pp \to 2 \Pa + \mathrm{X}) \, \mathcal{B}^2 (\Pa \rightarrow 2\mu) \, \alpha_{\text{gen}} = N(m_{\mu \mu})/\left(\mathcal{L}  \bar{r} \right)$, where $\alpha_{\text{gen}}$ is the generator-level kinematic and geometric acceptance defined earlier. The calculation uses the integrated luminosity $\mathcal{L} = 20.7\fbinv$ and central value $\bar{r}$ of the ratio $r = \epsilon_{\text{data}} / \alpha_{\text{gen}} = 0.63 \pm 0.07$. The ratio includes a scale factor correcting for experimental effects not included in the simulation and its uncertainty covers the variation in the ratio over all the benchmark model points used. The limit is calculated as a function of the dimuon mass using the $\text{CL}_{S}$ approach~\cite{Junk:1999kv, Read:2002hq}. The chosen test statistic is based on the profile likelihood ratio and is used to determine how signal- or background-like the data are. Systematic uncertainties are incorporated in the analysis via nuisance parameters with a log-normal probability density function and are treated according to the frequentist paradigm. The overall statistical methodology used in this analysis was developed by the ATLAS and CMS Collaborations in the context of the LHC Higgs Combination Group and is described in~\cite{ATLAS:2011tau, Chatrchyan:2013lba}. The obtained limit as a function of dimuon mass $m_{\mu\mu}$ can be conveniently approximated as a constant everywhere except the vicinity of the observed event, where it follows a Gaussian distribution:
\begin{equation*}
N(m_{\mu \mu}) \leq 3.1 + 1.2 \, \exp{ \left( - \frac{( m_{\mu\mu} - 0.32 )^2}{2 \times 0.03^2} \right) },
\end{equation*}
resulting in
\ifthenelse{\boolean{cms@external}}{
\begin{multline*}
\sigma(\Pp\Pp \to 2 \Pa + \mathrm{X}) \, \mathcal{B}^2 (\Pa \rightarrow 2\mu) \, \alpha_{\text{gen}}\\
\leq 0.24 + 0.09 \, \exp{ \left( - \frac{( m_{\mu\mu} - 0.32 )^2}{2 \times 0.03^2} \right) },
\end{multline*}
}{
\begin{equation*}
\sigma(\Pp\Pp \to 2 \Pa + \mathrm{X}) \, \mathcal{B}^2 (\Pa \rightarrow 2\mu) \, \alpha_{\text{gen}}
\leq 0.24 + 0.09 \, \exp{ \left( - \frac{( m_{\mu\mu} - 0.32 )^2}{2 \times 0.03^2} \right) },
\end{equation*}
}
where $m_{\mu\mu}$ is measured in \GeV and the cross-section limit is expressed in femtobarns. This limit is applicable to models with two pairs of muons coming from light bosons of the same type with a mass in the range $2 m_{\mu} < m_{\Pa} < 2 m_{\tau}$, where the new light bosons are typically isolated and spatially separated (so as to satisfy the isolation requirements).

The weak model dependence of the ratio $r$ allows for a simple reinterpretation of the results in other models. This requires calculating $\alpha_{\text{gen}}$, as defined earlier, and then the full efficiency $\epsilon_{\text{data}}$ can be calculated by multiplying $\alpha_{\text{gen}}$ by the ratio $r$.

There are certain subtleties that must be taken into account when reinterpreting the model-independent results of this analysis in the context of other models, particularly with the isolation requirement. An event should be considered to satisfy the selection requirements if there are at least two well isolated $\gamma_{\mathrm{D}}$ decaying to muon pairs. Experimentally, isolation is based on charged tracks but it may be insufficient to just require the absence of generator-level charged particles in the isolation cone. For example a neutral pion decaying to a pair of photons, that convert into electrons, may result in the reconstruction of one or more tracks. This would be particularly relevant for models with more than two dark photons produced in the same event, some of which may decay to hadrons or electrons. In this case the safest approach is to require that there are no particles with $\pt > 0.5$\GeV within the $\gamma_{\mathrm{D}}$ isolation cone. This restriction would result in a more conservative limit but it would be robust against these effects.

\begin{figure*}[tb]
\centering
\includegraphics[width=0.485\textwidth]{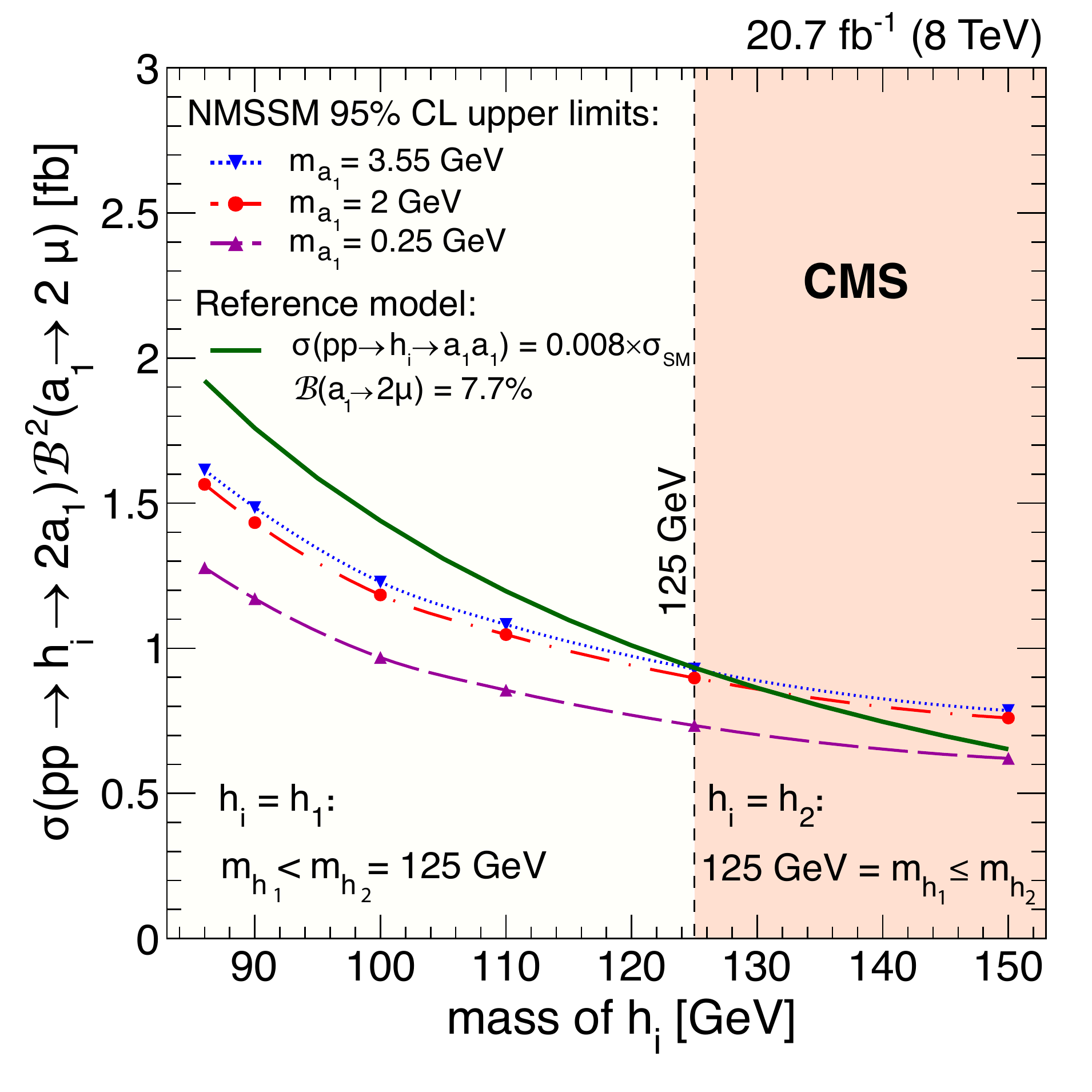}
\includegraphics[width=0.495\textwidth]{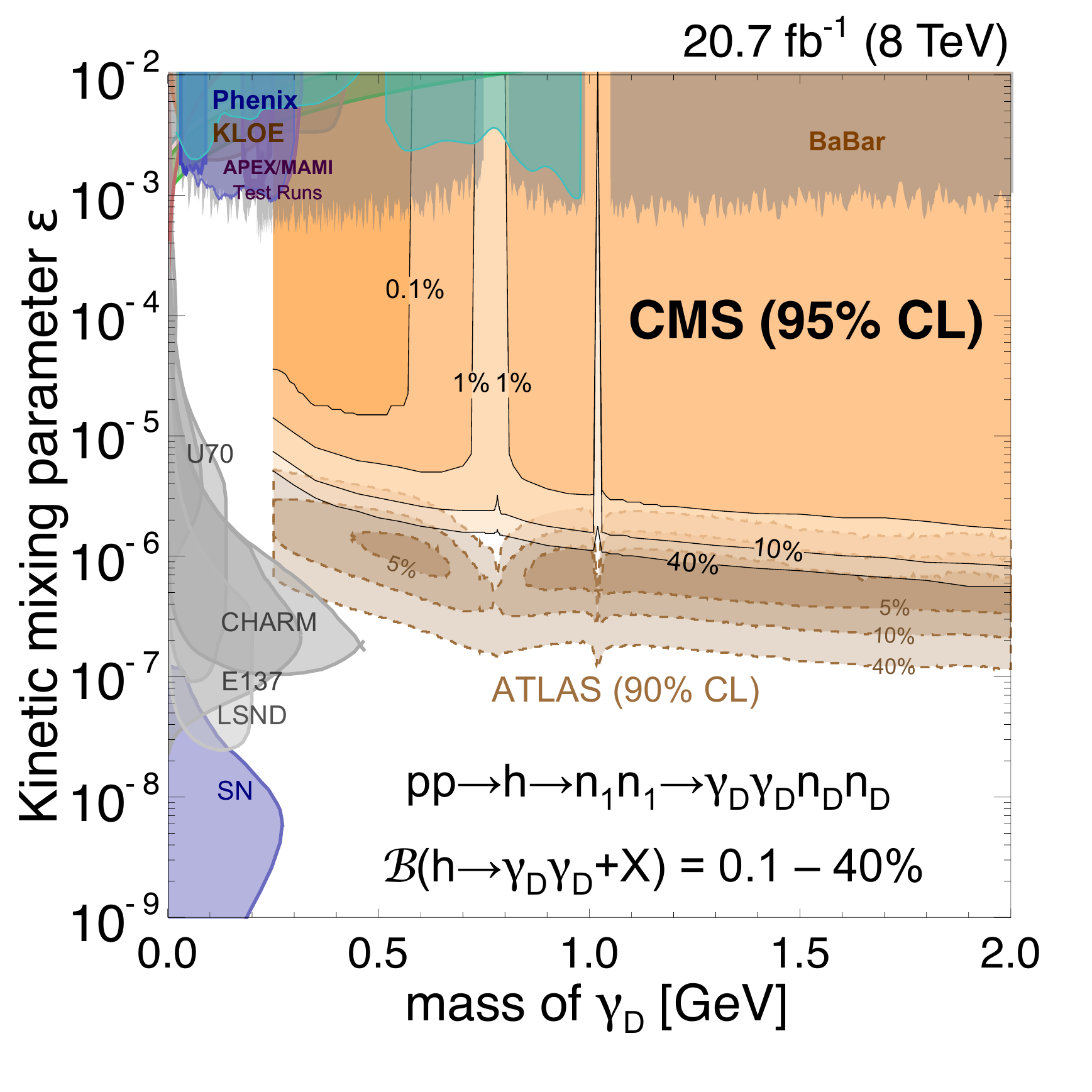}
\caption{Left for benchmark model 1: $95\%$ CL upper limits from this search for the NMSSM scenarios with $m_{\Pa_1}=0.25$\GeV (dashed curve), $m_{\Pa_1}=2$\GeV (dash-dotted curve) and $m_{\Pa_1}=3.55$\GeV (dotted curve) on $\sigma(\Pp\Pp \to \Ph_{1/2} \to 2 \Pa_1) \, \mathcal{B}^2(\Pa_1 \to 2 \mu)$ as a function of $m_{\Ph_{1}}$ in the range $86 < m_{\Ph_1} < 125 $\GeV and of $m_{\Ph_{2}}$ for $m_{\Ph_2} > 125 $\GeV. As an illustration, the limits are compared to the predicted rate (solid curve) obtained using a simplified scenario with $\sigma(\Pp\Pp \to \Ph_i \to 2a_1) = 0.008 \, \sigma_{\mathrm{SM}}$, which yields predictions for the rates of dimuon pair events comparable to the obtained experimental limits, and $\mathcal{B}(\Pa_1 \to 2\mu) = 7.7\%$. The chosen $\mathcal{B}(\Pa_1 \to 2\mu)$ is taken from~\cite{Dermisek:2010mg} for $m_{\Pa_1} = 2$\GeV and $\tan \beta = 20$. Right for benchmark model 2: 95\% CL upper limits (black solid curves) from this search on $\sigma(\Pp\Pp \to \Ph \to 2\gamma_{\mathrm{D}} + X) \, \mathcal{B}(\Ph \to 2\gamma_{\mathrm{D}} + X)$ (with $m_{\mathrm{n}_1}=10$\GeV, $m_{\mathrm{n}_{\mathrm{D}}}=1$\GeV) in the plane of two of the parameters ($\varepsilon$ and $m_{\gamma_{\mathrm{D}}}$) for the dark SUSY scenarios, along with constraints from other experiments~\cite{Aad:2014yea, Adare:2014mgk, Babusci:2014sta, Abrahamyan:2011gv, Merkel:2014avp, Agakishiev:2013fwl, Lees:2014xha, Fradette:2014sza,Essig:2013lka,2012arXiv1201.2683D,Dreiner:2013mua,Blumlein:2011mv,Essig:2010gu,Batell:2009di,Gninenko:2012eq} showing the 90\% CL exclusion contours. The colored contours represent different values of $\mathcal{B}(\Ph \to 2\gamma_{\mathrm{D}} + X)$ in the range 0.1--40\%.\label{fig:limits_combined}}
\end{figure*}

The results from this analysis are also interpreted in the context of the NMSSM and the dark SUSY benchmark models, and 95\% CL upper limits on the product of the cross section and branching fraction are derived. In these models both the Higgs boson production cross section and the branching fractions can vary significantly, depending on the choice of parameters. In the absence of broadly accepted benchmark scenarios, the production cross sections in these examples are normalized to that of the SM Higgs boson with a mass of 125\GeV~\cite{Dittmaier:2011ti}.

In the case of the NMSSM benchmark scenario, the production cross sections and branching fractions for h$_1$ and h$_2$ can vary substantially depending on the chosen parameters. An exact interpretation of these results requires evaluating the experimental acceptance using the generator-level acceptance for each of the two $\Ph_{1,2}$ bosons, and then using the measured upper limit on the sum of two contributions to derive limits for any choice of NMSSM model parameters. To present results in a fashion allowing for straightforward interpretation, we note that if one of the two CP-even Higgs bosons is the 125\GeV state observed at the LHC, then the other one is either lighter or heavier. In the NMSSM it is typical that one of the two has approximately the SM production cross section and a small $\mathcal{B}(\Ph_i\rightarrow 2a)$, whereas the other one has a suppressed production rate and large $\mathcal{B}(\Ph_i\rightarrow 2a)$ due to its large singlet fraction. In Fig.~\ref{fig:limits_combined} (left) the limit at each mass point is calculated taking the CP-even Higgs boson with the corresponding mass as the only source of signal events; the curve below 125\GeV applies to NMSSM models in which $m_{\Ph_1} < m_{\Ph_2} = 125$\GeV, with h$_1$ decays dominating the rate of $4\mu$ events. The limit at $m_h = 125$\GeV corresponds to the case where $125\GeV  = m_{\Ph_1} < m_{\Ph_2} $, with h$_1$ decays still responsible for the vast majority of signal-like events. The points above 125\GeV correspond to model points for which only h$_2$ ($m_{\Ph_2}>m_{\Ph_1}=125 $\GeV) is allowed to have a sizeable rate of observable $4\mu$ events. Finally, for models with $m_{\Ph_2}>150$\GeV, the limit at 150\GeV can be used as a conservative estimate of the production rate limit. In each of these scenarios it is possible that the other Higgs boson also contributes some fraction of the $4\mu$ signal events, in which case the limit shown is more conservative than would be given by an exact evaluation.

In the case of the dark SUSY scenario, a 95\% CL limit on the product of the Higgs boson production cross section and the branching fractions of the Higgs boson (cascade) decay to a pair of dark photons is determined. The limit set in the $(m_{\gamma_{\mathrm{D}}}, \varepsilon)$ plane from this analysis is shown in Fig.~\ref{fig:limits_combined} (right), along with limits from other experimental searches, where the lifetime is directly related to the kinetic mixing parameter $\varepsilon$ and the mass of the dark photon $m_{\gamma_{\mathrm{D}}}$ via $\tau_{\gamma_{\mathrm{D}}} (\varepsilon, m_{\gamma_{\mathrm{D}}}) = \varepsilon^{-2} \, f(m_{\gamma_{\mathrm{D}}})$, where $f(m_{\gamma_{\mathrm{D}}})$ is a function that depends only on the mass of the dark photon~\cite{Batell:2009yf}. The significant vertical structures in the limits visible in Fig.~\ref{fig:limits_combined} (right) arise because the total width of the dark photon varies rapidly in those mass regions due to resonant decays to hadrons. This search constrains a large, previously unconstrained area of the parameter space. Unlike the other results in the figure, the CMS and ATLAS limits are model-dependent and only valid under the assumption that $\mathcal{B}(\Ph\rightarrow2\mathrm{n}_1\rightarrow 4\mu+X)\neq0$. The recent ATLAS analysis~\cite{Aad:2014yea} focused on highly displaced objects and these searches therefore probe different regions of the available parameter space.

\section{Summary}

A search for pairs of new light bosons produced in the decay of a Higgs boson, that subsequently decay to pairs of oppositely charged muons, is presented. One event is observed in the signal region, with $2.2 \pm 0.7$ events expected from the SM backgrounds. A model independent upper limit at 95\% CL on the product of the cross section, branching fraction, and acceptance is obtained. This limit is valid for light boson masses in the range $2 m_{\mu} < m_{\Pa} < 2 m_{\tau}$. The obtained results allow a straightforward interpretation within a broad range of physics models that predict the same type of signature. The results are compared with two benchmark models in the context of the NMSSM and dark SUSY, including scenarios predicting a nonnegligible light boson lifetime.

\begin{acknowledgments}
We would like to thank J. Ruderman for his guidance with the theoretically motivated benchmark samples of dark SUSY. We congratulate our colleagues in the CERN accelerator departments for the excellent performance of the LHC and thank the technical and administrative staffs at CERN and at other CMS institutes for their contributions to the success of the CMS effort. In addition, we gratefully acknowledge the computing centres and personnel of the Worldwide LHC Computing Grid for delivering so effectively the computing infrastructure essential to our analyses. Finally, we acknowledge the enduring support for the construction and operation of the LHC and the CMS detector provided by the following funding agencies: BMWFW and FWF (Austria); FNRS and FWO (Belgium); CNPq, CAPES, FAPERJ, and FAPESP (Brazil); MES (Bulgaria); CERN; CAS, MoST, and NSFC (China); COLCIENCIAS (Colombia); MSES and CSF (Croatia); RPF (Cyprus); MoER, ERC IUT and ERDF (Estonia); Academy of Finland, MEC, and HIP (Finland); CEA and CNRS/IN2P3 (France); BMBF, DFG, and HGF (Germany); GSRT (Greece); OTKA and NIH (Hungary); DAE and DST (India); IPM (Iran); SFI (Ireland); INFN (Italy); MSIP and NRF (Republic of Korea); LAS (Lithuania); MOE and UM (Malaysia); CINVESTAV, CONACYT, SEP, and UASLP-FAI (Mexico); MBIE (New Zealand); PAEC (Pakistan); MSHE and NSC (Poland); FCT (Portugal); JINR (Dubna); MON, RosAtom, RAS and RFBR (Russia); MESTD (Serbia); SEIDI and CPAN (Spain); Swiss Funding Agencies (Switzerland); MST (Taipei); ThEPCenter, IPST, STAR and NSTDA (Thailand); TUBITAK and TAEK (Turkey); NASU and SFFR (Ukraine); STFC (United Kingdom); DOE and NSF (USA).

Individuals have received support from the Marie-Curie programme and the European Research Council and EPLANET (European Union); the Leventis Foundation; the A. P. Sloan Foundation; the Alexander von Humboldt Foundation; the Belgian Federal Science Policy Office; the Fonds pour la Formation \`a la Recherche dans l'Industrie et dans l'Agriculture (FRIA-Belgium); the Agentschap voor Innovatie door Wetenschap en Technologie (IWT-Belgium); the Ministry of Education, Youth and Sports (MEYS) of the Czech Republic; the Council of Science and Industrial Research, India; the HOMING PLUS programme of the Foundation for Polish Science, cofinanced from European Union, Regional Development Fund; the Compagnia di San Paolo (Torino); the Consorzio per la Fisica (Trieste); MIUR project 20108T4XTM (Italy); the Thalis and Aristeia programmes cofinanced by EU-ESF and the Greek NSRF; and the National Priorities Research Program by Qatar National Research Fund.
\end{acknowledgments}

\bibliography{auto_generated}
\cleardoublepage \appendix\section{The CMS Collaboration \label{app:collab}}\begin{sloppypar}\hyphenpenalty=5000\widowpenalty=500\clubpenalty=5000\textbf{Yerevan Physics Institute,  Yerevan,  Armenia}\\*[0pt]
V.~Khachatryan, A.M.~Sirunyan, A.~Tumasyan
\vskip\cmsinstskip
\textbf{Institut f\"{u}r Hochenergiephysik der OeAW,  Wien,  Austria}\\*[0pt]
W.~Adam, E.~Asilar, T.~Bergauer, J.~Brandstetter, E.~Brondolin, M.~Dragicevic, J.~Er\"{o}, M.~Flechl, M.~Friedl, R.~Fr\"{u}hwirth\cmsAuthorMark{1}, V.M.~Ghete, C.~Hartl, N.~H\"{o}rmann, J.~Hrubec, M.~Jeitler\cmsAuthorMark{1}, V.~Kn\"{u}nz, A.~K\"{o}nig, M.~Krammer\cmsAuthorMark{1}, I.~Kr\"{a}tschmer, D.~Liko, T.~Matsushita, I.~Mikulec, D.~Rabady\cmsAuthorMark{2}, B.~Rahbaran, H.~Rohringer, J.~Schieck\cmsAuthorMark{1}, R.~Sch\"{o}fbeck, J.~Strauss, W.~Treberer-Treberspurg, W.~Waltenberger, C.-E.~Wulz\cmsAuthorMark{1}
% ...
\vskip\cmsinstskip
% ...
1:~~Also at Vienna University of Technology, Vienna, Austria\\
2:~~Also at CERN, European Organization for Nuclear Research, Geneva, Switzerland\\
% ...

\end{sloppypar}
\end{document}